\documentclass[reprint,aps,pre,superscriptaddress]{revtex4-1}
\usepackage{graphicx}
\usepackage{dcolumn}
\usepackage{bm}
\usepackage{amsmath}
\usepackage{soul}
\usepackage{xcolor}

\usepackage{hyperref}
\hypersetup{
    colorlinks=true,
    linkcolor=black,
    filecolor=black,      
    urlcolor=blue,
    citecolor=black
}

\begin{document}

\title{Resonances between fundamental frequencies for lasers with large
delayed feedbacks}
\author{Anton V. Kovalev}
\email{avkovalev@niuitmo.ru}
\affiliation{ITMO University, Birzhevaya Liniya 14, 199034 Saint Petersburg,
Russia}
\author{Md Shariful Islam}
\affiliation{Georgia Tech-CNRS UMI 2958, Georgia Tech Lorraine, 2 Rue Marconi,
57070 Metz, France}
\affiliation{School of Electrical and Computer Engineering, Georgia Institute of
Technology, Atlanta,Georgia 30332-0250, USA}
\author{A. Locquet}
\affiliation{Georgia Tech-CNRS UMI 2958, Georgia Tech Lorraine, 2 Rue Marconi,
57070 Metz, France}
\affiliation{School of Electrical and Computer Engineering, Georgia Institute of
Technology, Atlanta,Georgia 30332-0250, USA}
\author{D.S. Citrin}
\affiliation{Georgia Tech-CNRS UMI 2958, Georgia Tech Lorraine, 2 Rue Marconi,
57070 Metz, France}
\affiliation{School of Electrical and Computer Engineering, Georgia Institute of
Technology, Atlanta,Georgia 30332-0250, USA}
\author{Evgeny A. Viktorov}
\affiliation{ITMO University, Birzhevaya Liniya 14, 199034 Saint Petersburg,
Russia}
\author{Thomas Erneux}
\affiliation{Optique Nonlin\'{e}aire Th\'{e}orique, Universit\'{e} Libre de Bruxelles,  
Campus Plaine C.P. 231, 1050 Bruxelles, Belgium}
\affiliation{ITMO University, Birzhevaya Liniya 14, 199034 Saint Petersburg,
Russia}
\date{\today}

\begin{abstract}
High-order frequency locking phenomena were recently observed using
semiconductor lasers subject to large delayed feedbacks \cite{OE,chaos}.
Specifically, the relaxation oscillation (RO) frequency\ and a harmonic of
the feedback-loop round-trip frequency coincided with the ratios 1:5 to 1:11. By
analyzing the rate equations for the dynamical degrees of freedom in a laser subject to a delayed optoelectronic
feedback, we show that the onset of a two-frequency train of pulses occurs
through two successive bifurcations.\ While the first bifurcation is a %
primary Hopf bifurcation to the ROs, a secondary Hopf bifurcation
leads to a two-frequency regime where a low frequency, proportional to
the inverse of the delay, is resonant with the RO frequency.\ \ We
derive an amplitude equation, valid near the first Hopf bifurcation
point, and numerically observe the frequency locking.\ \ We
mathematically explain this phenomenon by formulating a closed system
of ordinary differential equations from our amplitude equation. Our
findings motivate new experiments with particular attention to the first two
bifurcations.\ We observe experimentally (1) the frequency locking phenomenon as we pass
the secondary bifurcation point, and (2) the nearly constant slow
period as the two-frequency oscillations grow in amplitude. Our
results analytically confirm previous observations of frequency locking
phenomena for lasers subject to a delayed optical feedback.

A. V. Kovalev, M. S. Islam, A. Locquet, D. S. Citrin, E. A. Viktorov, and T. Erneux, \href{https://dx.doi.org/10.1103/PhysRevE.99.062219}{Phys. Rev. E \textbf{99}, 062219 }(\href{https://dx.doi.org/10.1103/PhysRevE.99.062219}{2019}).
\copyright{ 2019 American Physical Society}

\end{abstract}

\maketitle

\section{Introduction}

By contrast to solid state or gas lasers, semiconductor lasers (SLs) are
sensitive to optical feedback because of the low reflectivity of the
internal mirrors \cite{kane}. Optical feedback can be intentionally
implemented, e.g., by external gratings and mirrors widely used for
stabilization and controlled tuning of the emission wavelength.\ On the
other hand, unintentional external feedback can occur from optical elements
in fiber-coupled modules, such as micro-lenses or fiber ends. Depending on
the application, optical feedback is either considered as a nuisance that
needs to be handled or a virtue, enabling the control of the operating
properties of light sources. The diverse applications of SLs in our daily
life (long-distance telecommunication, environmental sensing, code-bar
reading at the supermarket, laser printers) has driven rapid developments in
both theoretical and experimental studies on delayed feedback lasers. Today,
fundamental properties of delay induced phenomena, such as the
synchronization of delay-coupled oscillators or square-wave oscillations for
large delays, are conveniently studied in the laboratory using lasers or
other optical devices \cite{erneux,soriano,laurent}.

For a SL subject to a delayed optical feedback two typical frequencies play
an outsized role in determining the dynamical properties. First, the
relaxation oscillation (RO) frequency $f_{RO}$ is the frequency of weakly
damped oscillations measured at the output of the solitary laser, in the
absence of any feedback. Second, $f_{delay}=\tau ^{-1}$ is the inverse of
the round-trip time for the light to go from the laser to the mirror, and
back to the laser. $f_{RO}$ and $f_{delay}$ typically range in the GHz and
MHz time scales. As early as 1969, Broom \cite{broom,broom2} reported on a
resonant interaction between these two frequencies. He stated that ``the
interaction would be strongest when $f_{RO}=nf_{delay}$ where $n$ is a small
integer'' \cite{broom}. The hypothesis of such resonant instabilities if the
delay is large was revived in 2001 and 2003 by Liu and coworkers \cite%
{liu1,liu2} who explored the response of a SL subject to an optoelectronic
feedback on the injection current. They obtained bifurcation diagrams where
``frequency-locked regimes'' appear as a result of a secondary bifurcation
from a branch of sustained RO oscillations. The ratio $f_{delay}/f_{RO}$ was
in the range 10--30. More recently, this question reappeared in a series of
experiments carried on a quantum dot laser subject to optical feedback and
partial filtering \cite{OE}. The delay was large and the ratio $%
f_{RO}/f_{delay}$ was close to an integer $n$. The authors reported on a
Hopf bifurcation to sustained RO oscillations followed by a secondary
bifurcation to quasi-periodic oscillations as the feedback parameter is
progressively increased. The envelope of the fast RO oscillations is no
longer constant but slowly oscillates at frequency $f_{delay}.$ The ratio $%
f_{RO}/f_{delay}=n$ exhibited an integer with $n$ ranging from $5$ to $11$
depending on the laser parameters. This work is further explored in Ref.~
\cite{chaos} for both quantum dot and quantum well SLs. Resonant effects between fundamental frequencies in a delayed feedback laser system may sufficiently stabilize the laser output as has been recently demonstrated in Ref. \cite{PTL}.

The simplest laser system operating with a delayed feedback is the laser
subject to an optoelectronic feedback on the injection current. By contrast
to a laser subject to an optical feedback from a distant mirror, the phase
of the laser field plays a passive role, and only the laser intensity needs
to be taken into account \ \cite{TE}. This ideal setting was already
considered in 1989 by Giacomelli et al. \cite{gia} who studied the Hopf
bifurcation instabilities both experimentally and theoretically in terms of
feedback gain, delay, and pump parameter. Their model equations are
equivalent to Eqs.~(\ref{R1}) and (\ref{R2}) below \footnote{%
The change of variables and parameters from (\cite{gia}) to Eqs.~(\ref{R1})
and (\ref{R2}) are: $X=\frac{2}{\alpha }I,$ $Y=1+2N,$ $t\rightarrow kt,$ $%
\gamma \rightarrow \frac{k}{\gamma },$ $P=\frac{\alpha a}{2},$ $\eta =B,$ $%
\tau \rightarrow k\tau$.}, and particular attention was devoted to the Hopf
bifurcation frequencies. We note from their largest delay case that the
product of the Hopf bifurcation frequency and the delay is close to a large
multiple of $2\pi $ \footnote{%
From the last line in Table 1 of \cite{gia}, we compute $\omega
_{H}\tau /(2\pi )\simeq \Omega \tau /(2\pi )=1.789\times 5.87=10.45$ which
is close to $n=10$.}. This is the case we are investigating in this paper.%

For a laser subject to an optoelectronic feedback, a Hopf bifurcation
from a steady state is not the only mechanism generating time-periodic
oscillations of the intensity. Isolated branches of periodic solutions of
higher amplitude may coexist with the Hopf bifurcation branch \cite{pieroux1}%
. As the delay is progressively increased, these isolated branches
reduce in amplitude. The large delay limit is clearly a singular limit of
which we may take advantage by modifying the classical weakly nonlinear
analysis for a Hopf bifurcation. Indeed, we realize that a relatively large
delay not only perturbs the fast evolution of a basic oscillator (here, the
ROs) but also the slow evolution of the amplitude of the
oscillations (here, the slow damping of the ROs). A new two-time
scale analysis was developed and led to a slow time amplitude equation where 
a slow time delay appears \cite{pieroux}. All periodic solutions of
the original laser problem are now steady state solutions of this amplitude
equation. Asymptotic theories based on the large delay limit has
become a topic of high interest among physicists and mathematicians. It is
worth mentioning that two distinct approaches are possible (see Appendix \ref%
{MS}), from which we have chosen the one allowing us to analyze high-order
locking phenomena.

Our main goal is to analyze the resonance locking effect between $%
f_{RO}$ and $f_{delay}$ when the delay is large. To this end, we determine
an amplitude equation that captures the primary Hopf bifurcation and 
a secondary Hopf bifurcation to a two-frequency oscillatory regime.\
While the first frequency is clearly the RO frequency, we demonstrate that
the second frequency is locked to the first one and remains nearly
constant as we pass the secondary bifurcation point. The fact that the
period remains nearly constant as the bifurcation parameter is
changed is unusual for a Hopf bifurcation problem. It reminds us the
generation of square-wave oscillations in nonlinear scalar DDEs, such as the
Ikeda equation \cite{ikeda}. For these problems, the period of the
oscillations remains close to twice the delay, although the extrema of the
oscillations are functions of the control parameter. Here, however, we are
not dealing with square-waves but rather with nearly harmonic
oscillations, and a different analysis is needed.

The plan of the paper is as follows. In Section \ref{formulation}, we %
formulate the laser equations and observe the appearance of two-frequency
oscillations. We note that the ratio of the large and small frequencies is
close to a large integer. These observations then motivate a weakly
nonlinear analysis where both the weak damping of the RO oscillations and
the large delay are taken into account. All mathematical details are
relegated to the Appendix \ref{supp} for clarity. We derived a slow time
amplitude equation which we investigate for two specific cases. In Section %
\ref{noRO}, we consider the simplest mathematically possible case where the
contribution of the RO damping rate is neglected. Its simplicity allows us
to determine analytically the primary and secondary Hopf bifurcations
points. We then consider in Section \ref{RO} the more realistic case where
the natural damping rate of the RO oscillations is non-zero. Hopf
bifurcations lead to stable branches of solutions of growing amplitude but
with a period that remains nearly constant. In Section \ref{period},
we explain this phenomenon by assuming that the slow-time delay is large and
that the period of the oscillations is close to twice this delay.
Experiments using a single mode laser subject to a delayed optoelectronic
feedback substantiate our results by showing that the slow time period
remains constant as the feedback rate is increased. The experiments are
detailed in Section \ref{experiments}.   

\section{Laser equations}

\label{formulation}

In dimensionless form, the laser rate equations for the intensity of the
laser field $I$ and the carrier density $N$ are given by \cite{TE}: 
\begin{align}
I^{\prime }& =2NI,  \label{R1} \\
\gamma N^{\prime }& =P+\eta I(t-\tau )-N-(1+2N)I,  \label{R2}
\end{align}%
where $P=\mathcal{O}(1)$ is the value of the pump parameter above threshold in the
absence of feedback ($\eta =0).$ $\gamma =\mathcal{O}(10^{3})$ is the ratio of the
carrier and photon lifetimes. $\eta <1$ and $\tau =\mathcal{O}(10^{3})$ represent the
gain and the delay of the optoelectronic feedback, respectively. Because of
the large $\gamma $ and large $\tau ,$ these equations are delicate to solve
numerically, as we expect solutions exhibiting different time scales. A
change of variable allows us to eliminate the large $\gamma $ parameter
multiplying $N^{\prime }$ and reduces the size of the effective delay. The
new equations are derived in the Appendix \ref{supp} and are
given by 
\begin{align}
y^{\prime }& =x(1+y),  \label{R5} \\
x^{\prime }& =-y+\eta (1+y(s-\theta ))-\varepsilon x\left[ 1+2P(1+y)\right] ,
\label{R6}
\end{align}%
where $x$ and $y$ represents deviations of $N$ and $I$ from their steady
state values. Primes now mean differentiation with respect to $s\equiv
\omega t$ where $\omega \equiv \sqrt{2P/\gamma }\ll1$ is the relaxation
oscillation frequency of the laser. The new parameters $\theta $ and $%
\varepsilon $ are defined by 
\begin{equation}
\theta \equiv \omega \tau \text{ and }\varepsilon \equiv \frac{\omega }{2P}%
\ll1.  \label{R6a}
\end{equation}%

Physically, Eqs.~(\ref{R5}) and (\ref{R6}) with $\eta =\varepsilon =0$
describe the laser's natural ROs. The term multiplying $\varepsilon $
contributes to the slow damping of the relaxation oscillations in the
absence of feedback. The term multiplying $\eta $ accounts for the delayed
feedback.

\begin{figure}[tbp]
\includegraphics[width=\linewidth]{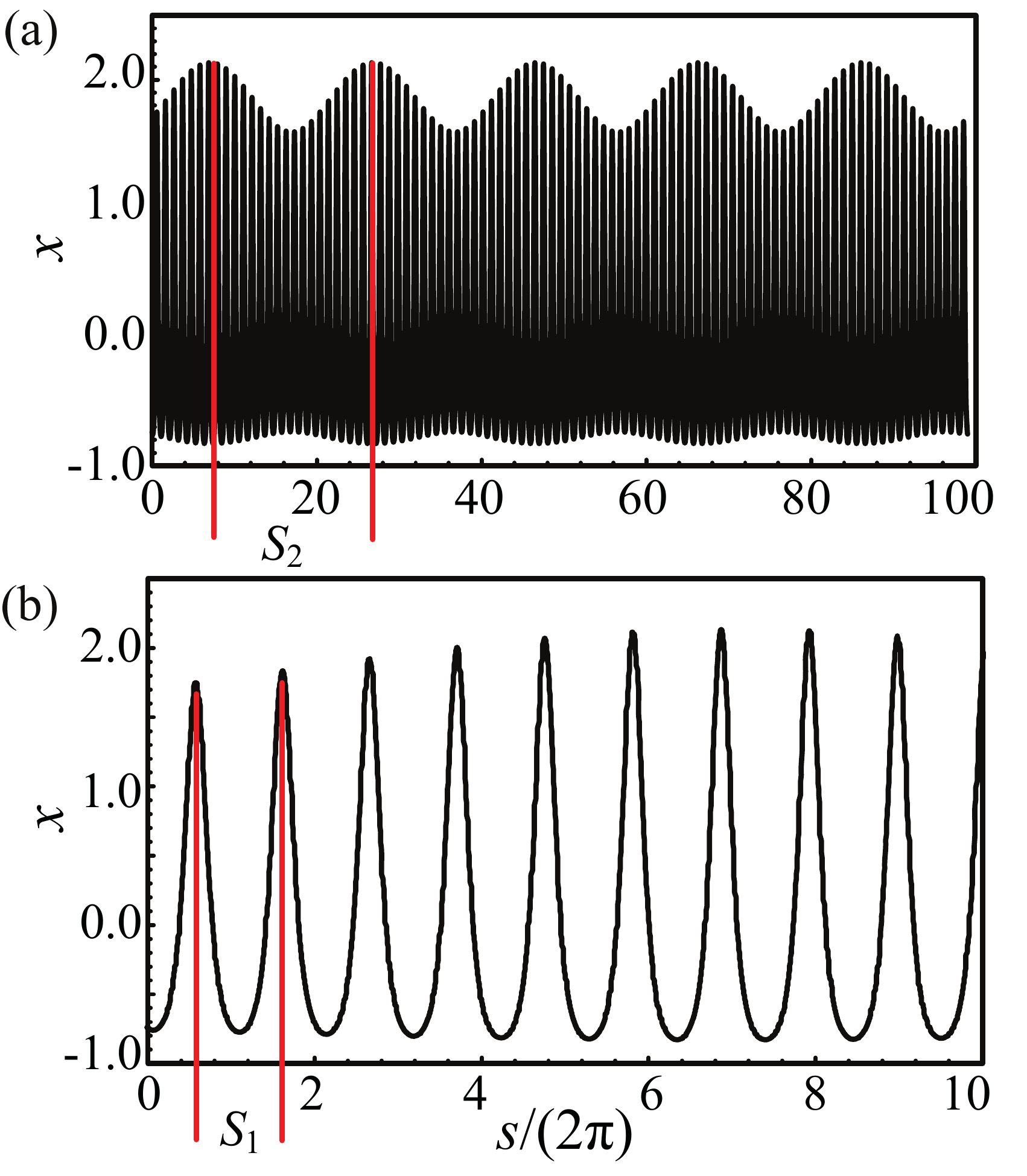}
\caption{Numerically obtained quasiperiodic oscillations. Both (a) and (b),
show deviation $x$ of the intensity $I$ from its steady state value in
different time scales for $\protect\eta =0.025,$ $\protect\theta =9.9\times 2%
\protect\pi $ and $\protect\varepsilon =0$. The two periods are $S_{1}=1.03$
and $S_{2}=19.81$.}
\label{eps1}
\end{figure}

Figure \ref{eps1} shows the long-time numerical solution of Eqs.~(\ref{R5})
and (\ref{R6}). The oscillations are quasiperiodic with two distinct
periods $S_{1}$ and $S_{2}$ (see Fig. \ref{eps1}) exhibiting a ratio $%
S_{1}/S_{2}=0.052\sim $~$1/20.$ They appear after a secondary bifurcation
point of a branch of $S_{1}$-periodic solutions. Simulations with
progressively higher $\theta \sim \theta _{n}=2n\pi $ suggest that this
secondary bifurcation point $\eta _{SB}$ verfies the scaling law 
\begin{equation}
\eta _{SB}\sim \frac{1}{n}\text{ (}n\rightarrow \infty ).  \label{R6c}
\end{equation}

\section{Weakly nonlinear analysis}

The fact that $\eta _{SB}$ $\rightarrow 0$ as $\theta \rightarrow \infty $
motivates a weakly nonlinear analysis where \ 
\begin{equation}
\delta \equiv 1/(2n)  \label{R52}
\end{equation}%
will be considered as a small parameter, and 
\begin{equation}
\eta =\delta b,\text{ }  \label{R52a}
\end{equation}%
where $b=\mathcal{O}(1).$ Furthermore, we scale the small parameter $\varepsilon $ in
a similar way as 
\begin{equation}
\varepsilon =\delta c,\text{ }  \label{R52b}
\end{equation}%
where $c=\mathcal{O}(1)$. The perturbation analysis is detailed in the Appendix \ref{supp}. We find that 
\begin{equation}
x=\delta ^{1/2}(iA(r)\exp (is)+c.c.)+\mathcal{O}(\delta ),  \label{R52d}
\end{equation}%
where $r\equiv \delta s$ is a slow time variable. The complex
amplitude $A(r)$ satisfies the following equation%
\begin{align}
2i\frac{dA}{dr}=\frac{1}{3}A^{2}A^{\ast }-Ab+bA(r-\delta \theta )\exp
(-i\theta )&  \notag \\
-ic(1+2P)A& ,  \label{R52c}
\end{align}%
which we now propose to explore. Eq.~(\ref{R52c}) was previously derived
(Eq.~(19) in \cite{pieroux}). Here, we concentrate on the quasiperiodic
oscillations of the laser equations which now correspond to periodic
solutions of Eq.~(\ref{R52c}). \ Of particular interest is the period of the
oscillations.

\subsection{No damping rate of the RO oscillations and perfect resonance}

\label{noRO}

We first examine the simple case $c=0$ and $\theta =\theta _{n}=2n\pi $. \
Introducing the decomposition $A=R\exp (i\phi )$ into Eq.~(\ref{R52c}), we
obtain from the real and imaginary parts 
\begin{align}
2R^{\prime }& =bR(r-\pi )\sin (\phi (r-\pi )-\phi ),  \label{A2} \\
2\phi ^{\prime }& =-\frac{R^{2}}{3}+b-b\frac{R(r-\pi )}{R}\cos (\phi (r-\pi
)-\phi ),  \label{A3}
\end{align}%
where prime now means differentiation with respect to $r$.\

\begin{figure}[tbp]
\includegraphics[width=\linewidth]{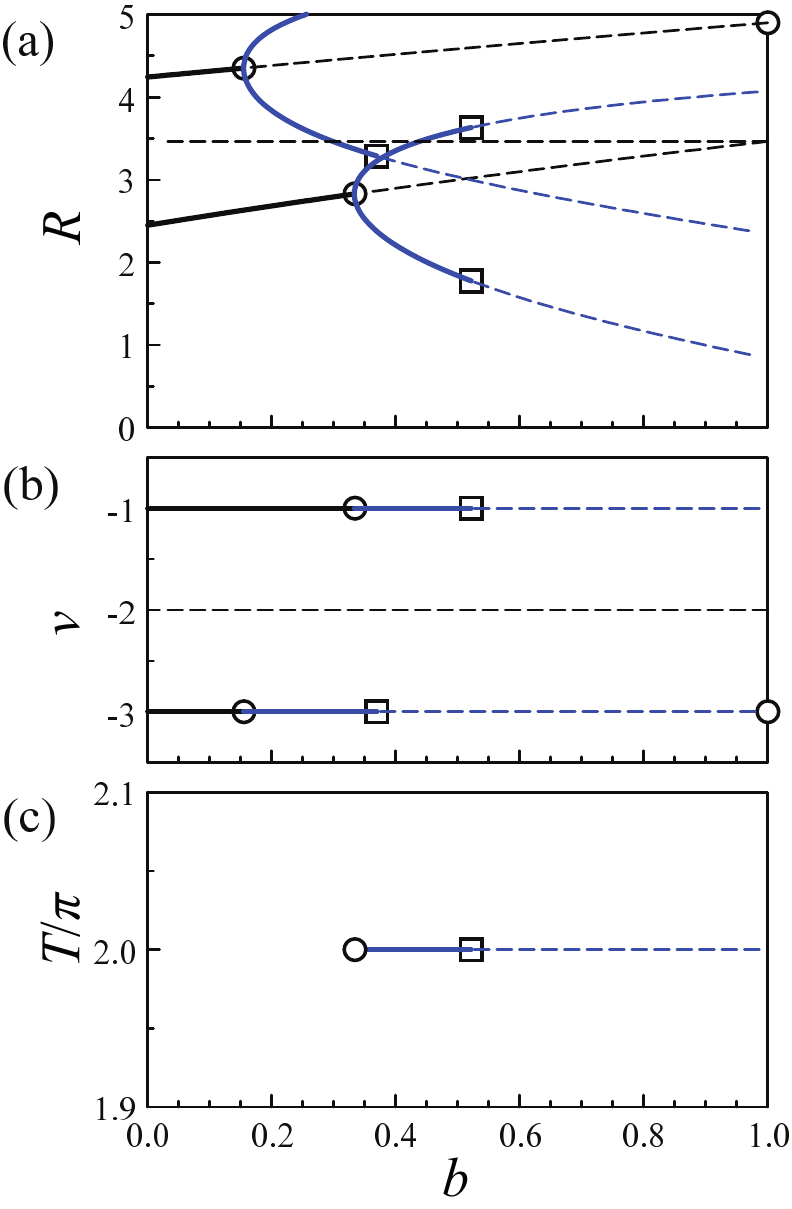}
\caption{Bifurcation diagram of the steady state (black) and periodic (blue)
solutions of the slow time amplitude equations  (\protect\ref{A2})
and (\protect\ref{A3}). We use the  decomposion $A=R\exp (i\protect%
\phi )$, where $\protect\phi =\protect\nu r$ is introduced in the text.
Figure (a) represents the extrema of $R$ as a function of the scaled
feedback strength $b$ for $\protect\theta =2n\protect\pi .$ They have been
obtained by using the numerical continuation method \protect\cite{biftool}.
Full and broken lines are stable and unstable solutions, respectively.\ \
The figure shows two stable branches of constant $R$ solutions emerging from 
$b=0.$ Circles are Hopf bifurcation points $b_{H}$ leading to stable
oscillations up to new bifurcation points $b_{PD}$ denoted by squares. \
Figure (b) represents the slow time frequency correction $\protect\nu $ for
the first three branches of steady states namely, $\protect\nu =-1,-2,$ and $%
-3$. Figure (c) shows the period of the periodic solution bifurcating from $%
R=\protect\sqrt{6(1+b)}$, $\nu = -1$ at $b_{H}=1/3$.\ The period remains constant as $b$
further increases from $b_{H}.$}
\label{BD1}
\end{figure}

These equations admit constant $R$ solutions with phase $\phi =\nu r$.\ In
terms of time $s$, the frequency of the basic RO oscillations in units
of the the orginal time $s$ now is $1+\delta \nu .\ $Fig. \ref{BD1}(a)
shows two stable and one unstable branch emerging from $b=0$.\ The two
stable branches depend on $b$ and are given by (see Appendix \ref{supp}) 
\begin{equation}
R=\sqrt{6(b-\nu )}\text{, and }\nu =-1,-3,  \label{A4}
\end{equation}%
while the unstable branch is independent of $b$ and admit the value 
\begin{equation}
R=\sqrt{12}.  \label{A4a}
\end{equation}%
The Hopf bifurcation point $b_{H}$ of Solution (\ref{A4}) with $\nu =-1$ is
determined analytically for arbitrary $\theta $ in the Appendix \ref{supp}.\ If $\theta =\theta _{n}$, it is located at 
\begin{equation}
b_{H}=\frac{1}{3}, \label{A5}
\end{equation}%
and agrees with the numerical estimate in Fig. \ref{BD1}.\ The Hopf
bifurcation frequency in units of time $r$ is equal to $1$ meaning that the
period of the oscillations at the bifurcation point equals $T=2\pi .$ Fig. %
\ref{BD1}(c) shows the period of the oscillations as their amplitude
increases ($b>$ $b_{H}).$\ Surprisingly, it remains close to $2\pi $.\ This
branch of periodic solutions changes stability at a new bifurcation point $%
b_{PD}$ (squares in Fig. \ref{BD1}(a)).\ Simulations indicate that it
corresponds to a period doubling bifurcation.\ 

\subsection{Non-zero RO damping rate and near resonant conditions}

\label{RO}

We now consider the case $c\neq 0$ and $\theta $ close, but different from $%
\theta _{n}=$ $2n\pi .\ $Introducing the decomposition $A=R\exp (i\phi )$
into Eq.~(\ref{R52c}), we obtain%
\begin{align}
2R^{\prime }& =bR(r-\delta \theta )\sin (-\theta +\phi (r-\delta \theta
)-\phi )-c(1+2P)R,  \label{R66} \\
2\phi ^{\prime }& =-\frac{1}{3}R^{2}+b-b\frac{R(r-\delta \theta )}{R}\cos
(-\theta +\phi (r-\delta \theta )-\phi ),  \label{R67}
\end{align}%
where prime means differentiation with respect to $r$.

\begin{figure}[tbp]
\includegraphics[width=\linewidth]{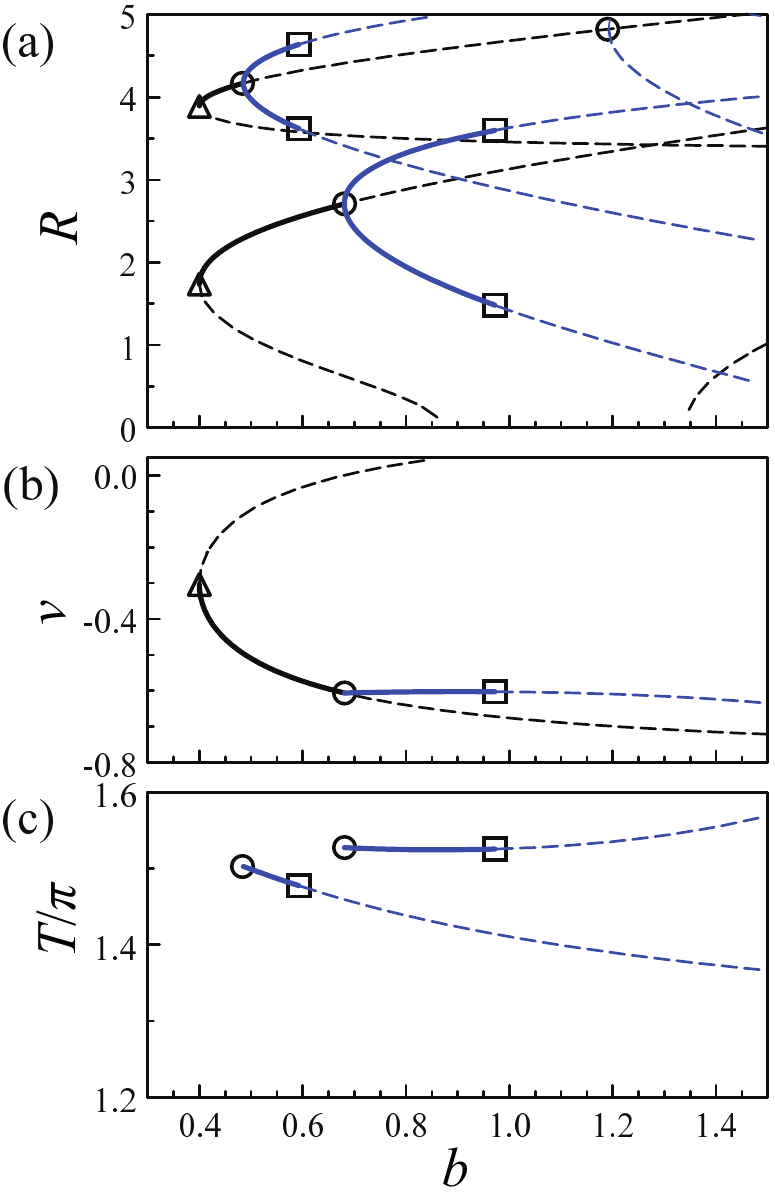}
\caption{Bifurcation diagram of the steady state and periodic solutions of
Eqs.~(\protect\ref{R66}) and (\protect\ref{R67}). Scaled decomposion $%
A=R\exp (i\protect\phi )$, where $\protect\phi =\protect\nu r$ is introduced
in the text. Figure (a) represents the extrema of $R$ as a function of the
scaled feedback strength $b.\ $The parameter values are $\protect\theta %
=9.9\times 2\protect\pi $ and $P=0.5.\ $ $\protect\theta $ is close to $%
\protect\theta _{n}=2n\protect\pi $ with $n=10.$ This then implies that $%
\protect\delta =1/(2n)=1/20,$ $c=0.2$ if $\protect\varepsilon =0.01$, and $%
\protect\delta \protect\theta =0.99\protect\pi .$ They have been obtained by
using the numerical continuation method \protect\cite{biftool}. The figure
shows two stable branches of constant $R$ solutions emerging from limit
points (triangles). The other notations and colors are the same as in Fig.~%
\protect\ref{BD1}. Figure (b) represents the slow time frequency correction $%
\protect\nu $ for the first branch$.$ Figure (c) shows the period of the
periodic solutions bifurcating from the two branches of constant $R$
solutions.\ Note that only the first Hopf bifurcation leads to a constant
period as we increase $b$.}
\label{BD2}
\end{figure}

The solutions
with $R=\text{const}$ and $\phi =\nu r,$ in parametric form, are given by ($\nu $ is
the parameter) 
\begin{align}
b& =-\frac{c(1+2P)}{\sin (\theta +\nu \delta \theta )},  \label{R68} \\
R^{2}& =3\left[ -2\nu +b-b\cos (\theta +\nu \delta \theta )\right] \geq 0.
\label{R69}
\end{align}%
Figure \ref{BD2} shows the bifurcation diagram for the extrema of $R$ as a
function of $b.$ The figure exhibits two Hopf bifurcations from two distinct
branches of constant $R$ solutions (two left circles in Fig. \ref{BD2}).\
Both bifurcations are leading to stable oscillations which become unstable
at new bifurcation points (squares in Fig. \ref{BD2}).

\begin{figure}[tbp]
\includegraphics[width=\linewidth]{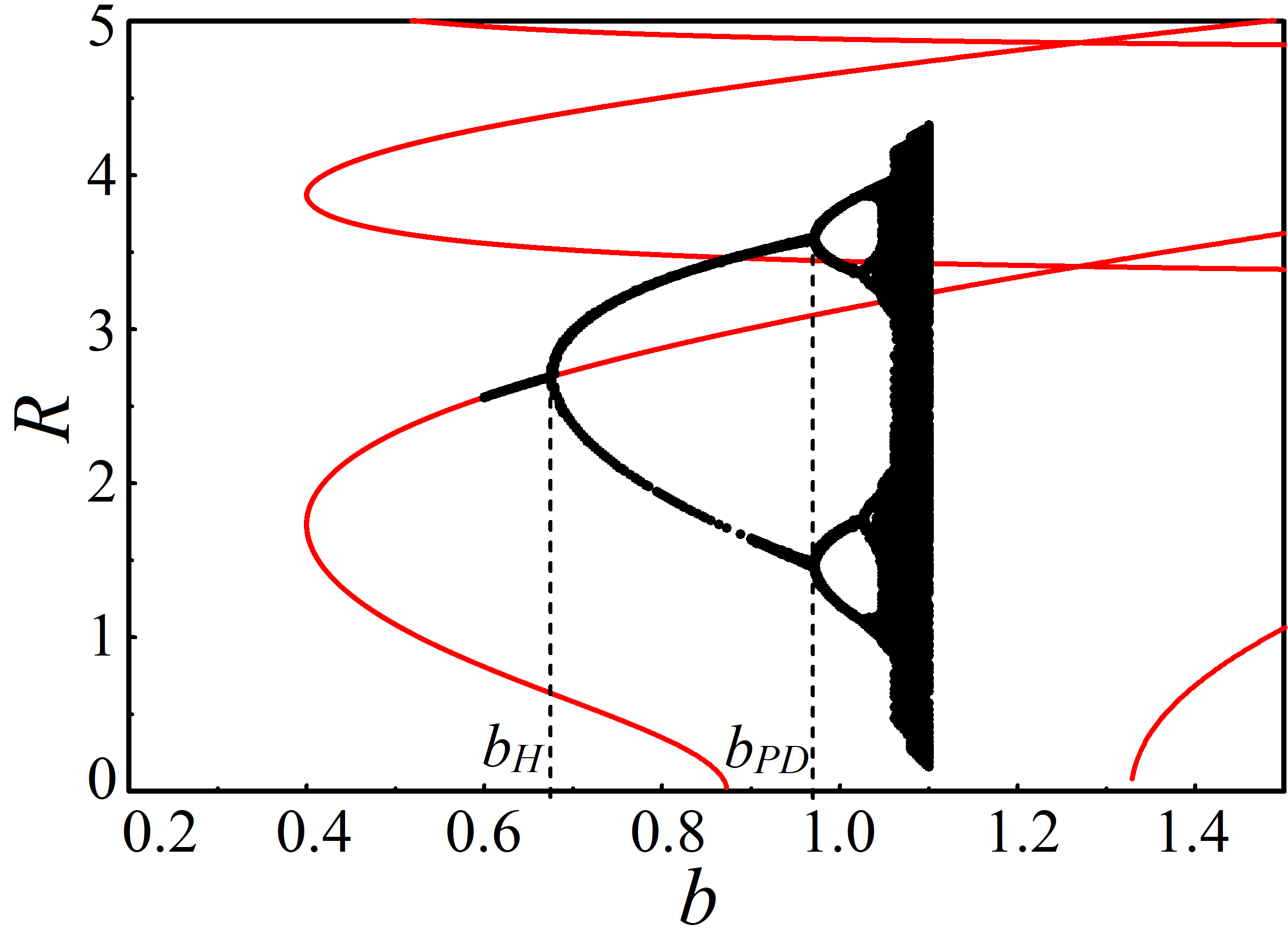}
\caption{Numerical bifurcation diagram of the extrema of $R$ as a function
of the scaled feedback strength $b$ for $\protect\theta =2n\protect\pi .$\
Scaled decomposion $A=R\exp (i\protect\phi )$, where $\protect\phi =\protect%
\nu r$ is introduced in the text. Long-time simulations were obtained from
Eq.~(\protect\ref{R52c}) reformulated in terms of the real and imaginary
parts of $A$.\ The original values of the parameters are: $\protect%
\varepsilon =10^{-2},$ $\protect\theta =9.9\times 2\protect\pi ,$ and $%
P=0.5.\ $ $\protect\theta $ is close to $\protect\theta _{n}=2n\protect\pi $
with $n=10$ implifying $\protect\delta =1/20,$ $c=0.2$, and $\protect\delta 
\protect\theta =0.99\protect\pi .$ The red lines are the constant $R$
solutions given by Eqs.~(\protect\ref{R68}) and (\protect\ref{R69}).\ }
\label{hopfbd1}
\end{figure}

Simulations of the amplitude Eq.~(\ref{R52c}), reformulated in terms of $%
A=u+iv$, for long intervals of time indicate that the secondary bifurcation
is a period doubling bifurcation and is followed by higher order
instabilities (Fig.\ \ref{hopfbd1}; only the bifurcation diagram
corresponding to the first Hopf bifurcation branch is shown for clarity).

We observe that the period of the first Hopf bifurcation branch
remains constant as the amplitude of the oscillations increases. The
period is no more nearly equal to $2\pi $ but is given by 
\begin{equation}
T/\pi =1.525\text{ }(b_{H}<b<b_{PD}).  \label{R70}
\end{equation}

\begin{figure}[tbp]
\includegraphics[width=\linewidth]{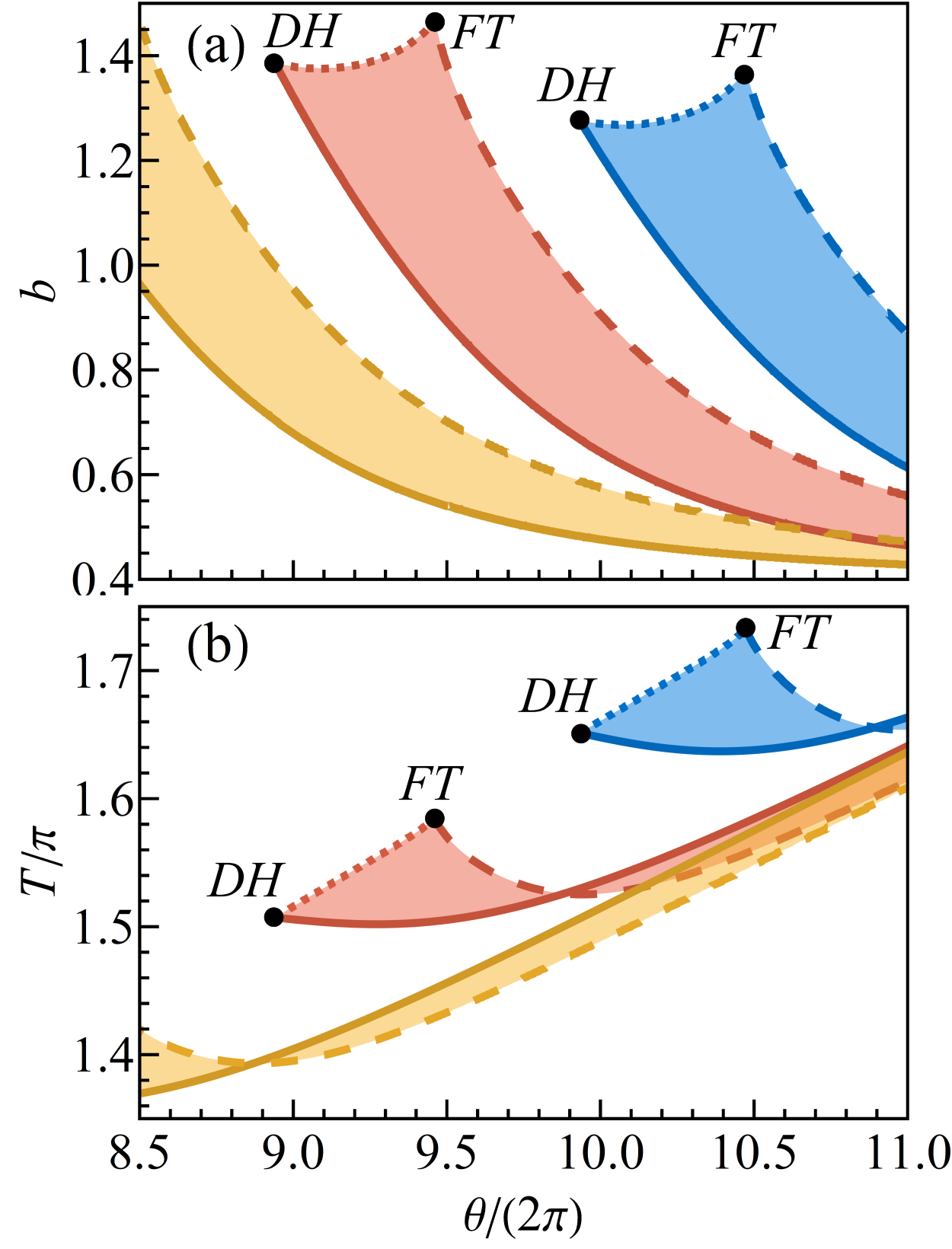}
\caption{Figure (a): domains of stable periodic solutions of Eqs.~(\protect
\ref{R66}) and (\protect\ref{R67}) in the $(\protect\theta ,b)$ plane;
figure (b): corresponding oscillation periods. Solid lines are the Hopf bifurcation
lines  leading to the periodic solutions. Dashed (dotted) lines stand for
period-doubling (torus) bifurcation lines. The colored areas delimit the
domains of stable oscillations. The red domain  is for the periodic
solutions that bifurcate from the first steady state branch if $\protect%
\theta _{c1}<\protect\theta <\protect\theta _{c2}$. The other colors
correspond to periodic solutions bifurcating from other neighbouring branches of steady
states. The black dots denote codimension-2 bifurcations located at the cusps: $DH$ are the double Hopf bifurcation points of the zero solution at $\theta _{c1}$ and $\theta _{c2}$; $FT$ are the flip-torus bifurcation points of the periodic solutions. The parameters are the same as in Fig.~\protect\ref{BD2}.}
\label{BD3}
\end{figure}

Accurate two parameter studies for $\theta $ close to $\theta
~=~10\times 2\pi $ have been determined by using a continuation
method, and are shown in Fig. \ref{BD3}. In Fig. \ref{BD3}(a) the stable
oscillations of Eqs.~(\ref{R66}) and (\ref{R67}) are bounded in the $b$ 
 versus $\theta$ plane by Hopf bifurcation lines (solid),
period-doubling bifurcation lines (dashed), and torus bifurcation lines
(dotted).\ The torus bifurcation leads to quasiperiodic oscillations. In this figure, the red delimits the
domain of stable periodic solutions connected to the first steady state
branch that bifurcates from zero if $\theta _{c1}<\theta <\theta _{c2}$ ($%
\theta _{c1}=8.94\times 2\pi $ and $\theta _{c2}=9.93\times 2\pi$, see 
Fig.~\ref{hopfrostab1} of the Appendix \ref{supp} for the primary Hopf bifurcation lines).\ 
At the double Hopf bifurcation point $\theta =\theta _{c2}$ two distinct steady state branches emerge
from zero at the same value of $b$. If $\theta
>\theta _{c2}$, a new steady state branch becomes the first to appear
from zero.\ The domain of stable oscillations
for this new branch is indicated in blue in Fig. \ref{BD3}.\ A similar
bifurcation scenario where a new steady-state branch becomes first occurs if  
$\theta <\theta _{c1}$.\ Its domain of stable oscillations is 
shown in orange in Fig.~\ref{BD3}. 

Figure~\ref{BD3}(b) shows the period as a function of $\theta$. At the intersections of the
full and dashed lines, the period is constant for the whole range of $b$ where the corresponding periodic solutions are stable. The bifurcation diagram shown in Fig.~\ref{BD2} is for $\protect\theta =9.9\times 2\protect\pi$ which corresponds to the intersection point of the red lines in Fig. 5. The periodic solution which stability domain is bounded by the red area in Fig.~\ref{BD3} exhibits the constant period as shown by
the upper line in Fig.~\ref{BD2}(c).

\section{Constant period}

\label{period}

In this section, we reconsider Eqs.~(\ref{A2}) and (\ref{A3}), which are
written for the scaled deviation of intensity $I$ from their steady state
values and plan to explain why the period remains constant as we pass the
Hopf bifurcation point.\ Our approach is similar to the analysis of time
periodic square-wave solutions of scalar delay differential equations
(DDEs).\ In Eqs.~(\ref{A2}) and (\ref{A3}), the delay equals $\pi $ and will
be treated as a large parameter.

We first introduce the variables $R_{j}$ and $\phi _{j}$ $(j=0,1,2)$ defined
by 
\begin{align}
R_{0}& \equiv R(r-\frac{\pi }{2}),\text{ }R_{1}\equiv R(r),\text{ }%
R_{2}\equiv R(r+\frac{\pi }{2}),  \label{R71} \\
\phi _{0}& \equiv \phi (r-\frac{\pi }{2}),\text{ }\phi _{1}\equiv \phi (r),%
\text{ and }\phi _{2}\equiv \phi (r+\frac{\pi }{2}).  \label{R72}
\end{align}%

We next consider two successive iterations of Eqs.~(\ref{A2}) and (\ref{A3}%
), namely%
\begin{align}
2R_{1}^{\prime }& =bR_{0}\sin (\phi _{0}-\phi _{1}),  \label{R73} \\
2R_{2}^{\prime }& =bR_{1}\sin (\phi _{1}-\phi _{2}),  \label{R74} \\
2\phi _{1}^{\prime }& =-\frac{R_{1}^{2}}{3}+b-b\frac{R_{0}}{R_{1}}\cos (\phi
_{0}-\phi _{1}),  \label{R75} \\
2\phi _{2}^{\prime }& =-\frac{R_{2}^{2}}{3}+b-b\frac{R_{1}}{R_{2}}\cos (\phi
_{1}-\phi _{2}).  \label{R76}
\end{align}%

The numerical simulations indicate that the period $p$ of the oscillations
is close to $2\pi .$ We therefore write 
\begin{equation}
p=2\pi +\alpha \text{ }  \label{R77}
\end{equation}%
where the correction $\alpha $ is assumed small compared to $2\pi .\ $With (%
\ref{R77}), the variables $R_{0}$ and $R_{1}$ are expanded as 
\begin{align}
R_{0}& \equiv R(r-\pi )=R(r-\frac{p}{2}+\frac{\alpha }{2})=R(r-\frac{p}{2}%
)+\mathcal{O}(\alpha ),  \label{R78} \\
R_{2}& \equiv R(r+\pi )=R(r+\frac{p}{2}-\frac{\alpha }{2})=R(r+\frac{p}{2}%
)+\mathcal{O}(\alpha ).  \label{R79}
\end{align}%

The periodicity condition now implies that 
\begin{equation}
R_{2}=R_{0}  \label{R80}
\end{equation}%
in first approximation.\ Similarly 
\begin{equation}
\phi _{2}=\phi _{0}.  \label{R81}
\end{equation}%
With (\ref{R80}) and (\ref{R81}), Eqs.~(\ref{R73}) and (\ref{R76}) reduce to
four ordinary differential equations 
\begin{align}
2R_{1}^{\prime }& =bR_{0}\sin (\phi _{0}-\phi _{1}),  \label{R82} \\
2R_{0}^{\prime }& =bR_{1}\sin (\phi _{1}-\phi _{0}),  \label{R83} \\
2\phi _{1}^{\prime }& =-\frac{R_{1}^{2}}{3}+b-b\frac{R_{0}}{R_{1}}\cos (\phi
_{0}-\phi _{1}),  \label{R84} \\
2\phi _{0}^{\prime }& =-\frac{R_{0}^{2}}{3}+b-b\frac{R_{1}}{R_{0}}\cos (\phi
_{1}-\phi _{0}).  \label{R85}
\end{align}%
Introducing $\Phi \equiv \phi _{1}-\phi _{0},$ we may eliminate one equation%
\begin{align}
2R_{1}^{\prime }& =-bR_{0}\sin (\Phi ),  \label{R86} \\
2R_{0}^{\prime }& =bR_{1}\sin (\Phi ),  \label{R87} \\
2\Phi ^{\prime }& =-\frac{R_{1}^{2}-R_{0}^{2}}{3}-b(\frac{R_{0}}{R_{1}}-%
\frac{R_{1}}{R_{0}})\cos (\Phi ).  \label{R88}
\end{align}%
From Eqs.~(\ref{R86}) and (\ref{R87}), we note a conservation relation given
by 
\begin{equation}
R_{0}^{2}+R_{1}^{2}=E,  \label{R89}
\end{equation}%
where $E$ is a positive constant.\ Solving numerically Eqs.~(\ref{A2}) and (%
\ref{A3}) for $b=0.5$, we find that $E(r)\equiv R^{2}(r)+R^{2}(r-\pi )$
oscillates close to a constant:%
\begin{equation}
E(r)=16.32\pm 0.02.  \label{R89a}
\end{equation}%
Using $R_{1}=\sqrt{E-R_{0}^{2}},$ we may further eliminate one equation and
obtain 
\begin{align}
2R_{0}^{\prime }& =b\sqrt{E-R_{0}^{2}}\sin (\Phi ),  \label{R90} \\
2\Phi ^{\prime }& =(2R_{0}^{2}-E)\left[ \frac{1}{3}-\frac{b}{R_{0}\sqrt{%
E-R_{0}^{2}}}\cos (\Phi )\right] .  \label{R91}
\end{align}%
One steady state is given by 
\begin{equation}
\Phi =\pi \text{ and }R_{0}^{2}=E/2.  \label{R91a}
\end{equation}%
From the linearized equation, we determine the characteristic equation for
the growth rate $\lambda $ 
\begin{equation}
4\left[ \lambda ^{2}+b(\frac{R_{0}^{2}}{3}+b)\right] =0.  \label{R93}
\end{equation}%
The $2\pi $ periodicity condition requires that $\lambda =i$ and (\ref{R93})
simplifies as 
\begin{equation}
-1+b(\frac{R_{0}^{2}}{3}+b)=0.  \label{R94}
\end{equation}%

We have verified that the expression of the steady state Eq.~(\ref{A4}) and
its bifurcation point Eq.~(\ref{A5}) identically satisfy Eq.~(\ref{R94}). We
conclude that Eqs.~(\ref{R90}) and (\ref{R91}) correctly predict the
previously determined Hopf bifurcation point. By dividing Eqs.~(\ref{R90})
and (\ref{R91}), we obtain a first order equation for $\cos (\Phi )$ as a
function of $R_{0}.$ This equation can be integrated and its solution
exhibits a new constant of integration $C$.

\begin{figure}[tbp]
\includegraphics[width=\linewidth]{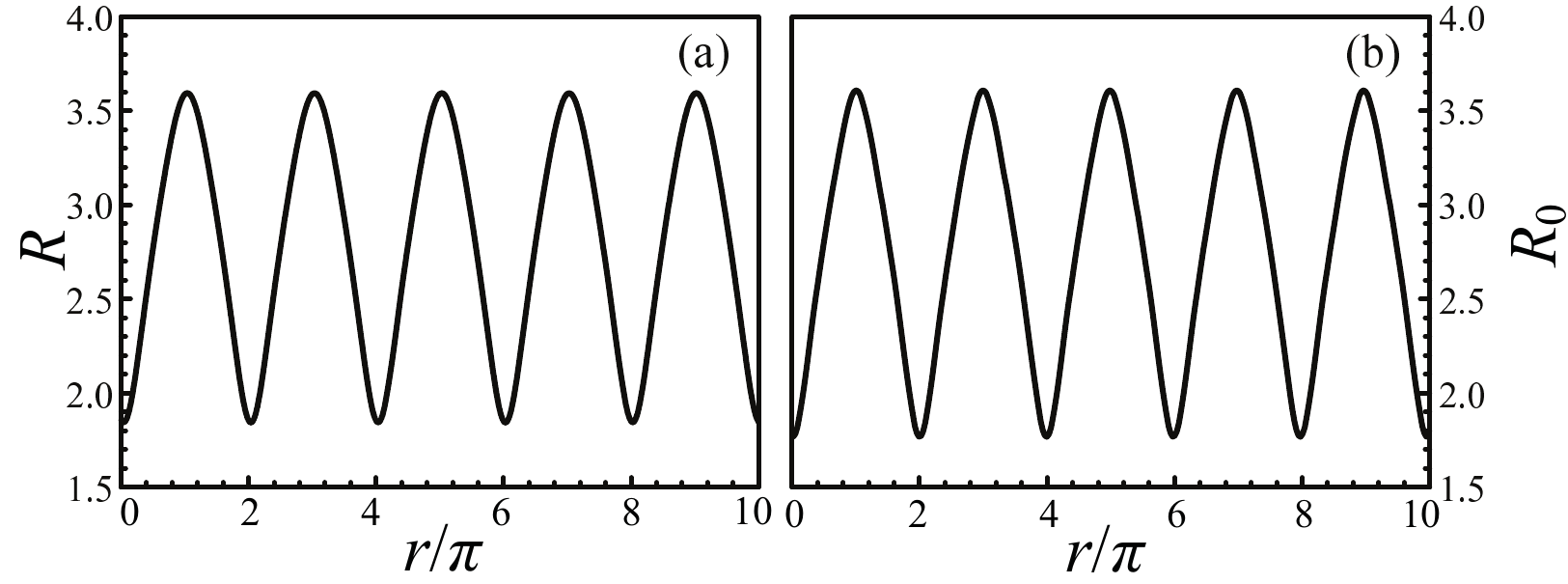}
\caption{(a): Numerical solution of the DDEs (\protect\ref{A2}) and (\protect
\ref{A3}). Scaled decomposion $A=R\exp (i\protect\phi )$, where $\protect%
\phi =\protect\nu r$ is introduced in the text. (b): Numerical solution of
the ODEs (\protect\ref{R90}) and (\protect\ref{R91}) with $E=16.32$, and
initial conditions $R_{0}(0)=3.61$, $\Phi (0)=\protect\pi .$ Parameter $%
b=0.5.$\ }
\label{optocomp}
\end{figure}

In summary, the analysis of the leading order equations indicates that the
amplitude and period of the oscillations depend on the values of two unknown
constants $E$ and $C$. Therefore, we need to explore higher order problems
and formulate two solvability conditions with respect to $E$ and $C$. The
higher order problems will exhibit the correction of the frequency $\alpha $
and we need the third condition. It is provided by the periodicity condition
of $R_{0}$ and $\phi _{0}.$ The higher order analysis is beyond the scope of
this paper.\ Our main objective was the derivation of the ODEs Eqs.~(\ref{R90}%
) and (\ref{R91}) from the original DDE problem Eqs.~(\ref{A2}) and (\ref{A3}%
). In order to substantiate our analysis, we have arbitrary fixed the
parameters $E$ and $C$ and solved Eqs.~(\ref{R90}) and (\ref{R91}) for $%
b=0.5 $ with the goal of finding the best fit to the numerical solution of
the full Eqs.~(\ref{A2}) and (\ref{A3}).\ The value of $E=16.32$ is
motivated by Eq.~(\ref{R89a}), and the value of $C$ is determined by choosing
the initial conditions. Since the maximum of $R$ appears when $\Phi =\pi ,$
we consider $\Phi (0)=\pi $ and only modify $R(0)$ so that the period of the
oscillations equals $2\pi .$ Fig. \ref{optocomp} compares the time traces of
the original DDEs and reduced ODEs. The agreement is excellent.

\section{Experiments}

\label{experiments}

Our mathematical analysis considered the rate equations for a
semiconductor laser subject to a delayed optoelectronic feedback and
predicted a resonance locking effect between the RO laser frequency and a
much lower frequency appearing through a secondary bifurcation mechanism.\
The latter is inversely proportional to the delay and exhibits a value that
remains nearly constant as we increase the control parameter.\ This
unexpected property from a bifurcation theory point of view motivates our
experiments.\ 

The device used for the experiments is a single mode
edge-emitting distributed feedback (DFB) p-doped InAs/InP QDash laser with a
cavity length of \ 500 $\mu $m, operating at 1550 nm. The DFB laser used has
an active region consisting of a stack of six layers of InAs quantum dashes,
each layer being embedded within an InGaAsP quantum well and separated by
InGaAsP barriers, AR/AR coated facets and a threshold current of 33 mA at
room temperature. The side mode suppression ratio was in excess of 40 dB in
the whole range of pumping used in the experiments.

\begin{figure}[tbp]
\includegraphics[width=\linewidth]{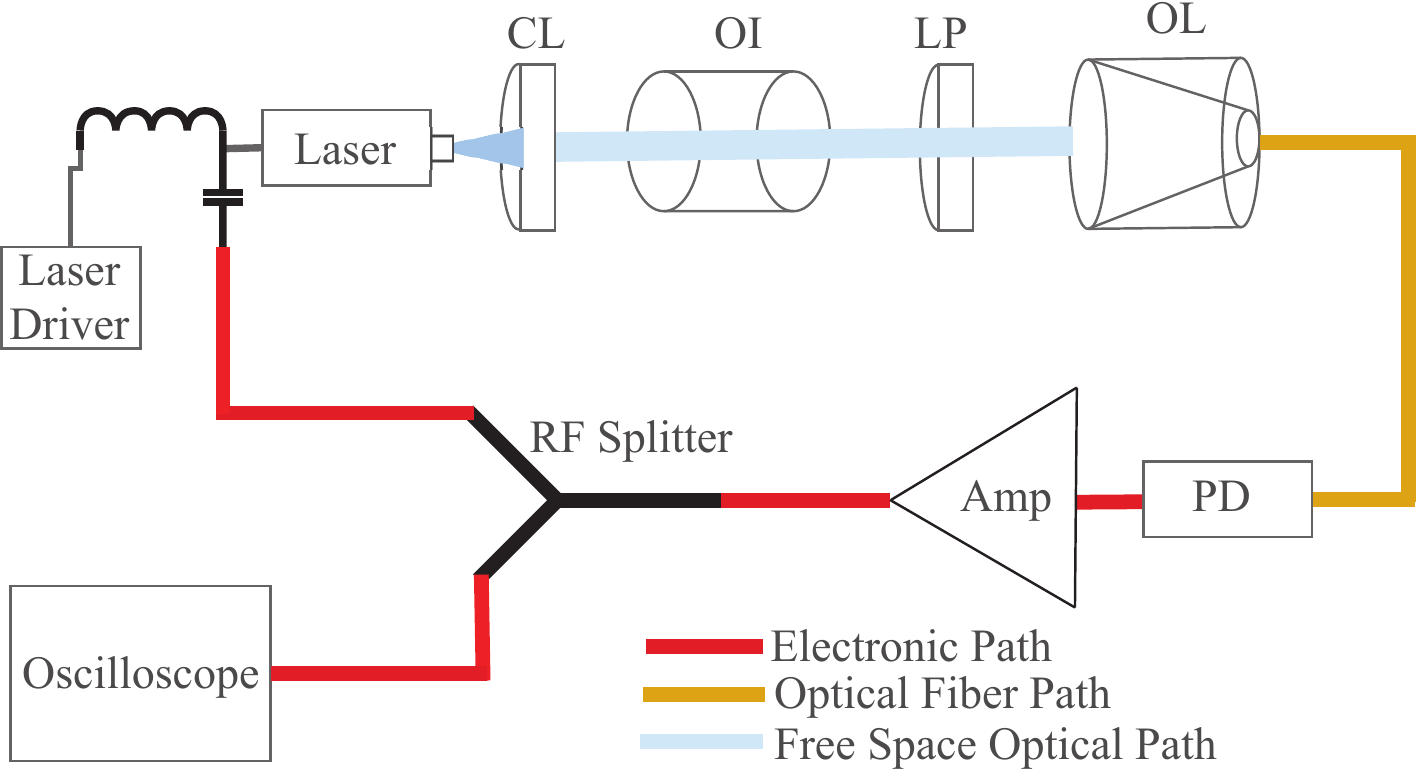}
\caption{Experimental arrangement of a laser diode subjected to
optoelectronic feedback. The optical intensity is photodetected, amplified
and added to the DC bias current of the laser diode. An oscilloscope
monitors the amplifier output. Various feedback levels are obtained by
rotating the linear polarizer in the free space optical path. CL --
collimating lens, OI -- optical isolator, LP -- linear polarizer, OL --
objective lens, PD -- photodetector.}
\label{setup}
\end{figure}

 The experimental setup
is shown in Fig.~\ref{setup}. The OE feedback consists of three stages. The
first stage corresponds to 35 cm of free space optical path which provides %
$\sim$1.17 ns delay. This free space path includes a collimating lens (CL),
an optical isolator (OI) that prevents back reflections into the laser, a
linear polarizer (LP) and an objective lens (OL) that focuses the light onto
an optical fiber. The second stage corresponds to 35~cm of optical patch
cable, providing $\sim$1.73 ns delay, and leading to a high-bandwidth
photodetector (12 GHz Newfocus 1544-B). The third stage is the electronic
path which starts with the photodetector and whose output is amplified
before being fed back to the laser. The amplification is implemented by
cascading a 18 dB amplifier with 20 GHz bandwidth (Newport 1422-LF) and a 30
dB amplifier with 30 GHz bandwidth (Microsemi UA0L30VM). The delay of the
electronic path, which also includes 70 cm of microwave coaxial cable and a
high frequency splitter (Mini-circuits ZX-10-2-183-S+), is measured to be 
$\sim$5.67 ns. The total delay of the OE feedback loop is then
estimated as $1.17+1.73+5.67=8.57$ ns ($\pm 0.20$~ns). The experimental
effective feedback level $\eta $ is here a relative measure of the feedback
strength and cannot be directly compared to the theoretical $\eta $ used in
our analysis. The parameter $\eta $ is controlled through the linear
polarizer (LP); it is affected by the responsivity of the photodetector and
the gain of the two amplification stages. Conventionally, $\eta =1$
corresponds to full transmission by the linear polarizer.\ Rotating the LP
allows a nonlinear control of the feedback level, since the attenuation
introduced varies as the cosine square of the angle between the LP and the
direction of polarization of the laser. Finally, a 50/50 RF splitter (18
GHz) was used to simultaneously feed the laser back and to monitor the
signal with the oscilloscope (12 GHz bandwidth, Agilent DSO80804B). The
splitter output used for feedback was directly added to the DC bias current
of the laser through a bias tee (26.5 GHz Marki BT-0026).

\begin{figure}[tbp]
\includegraphics[width=\linewidth]{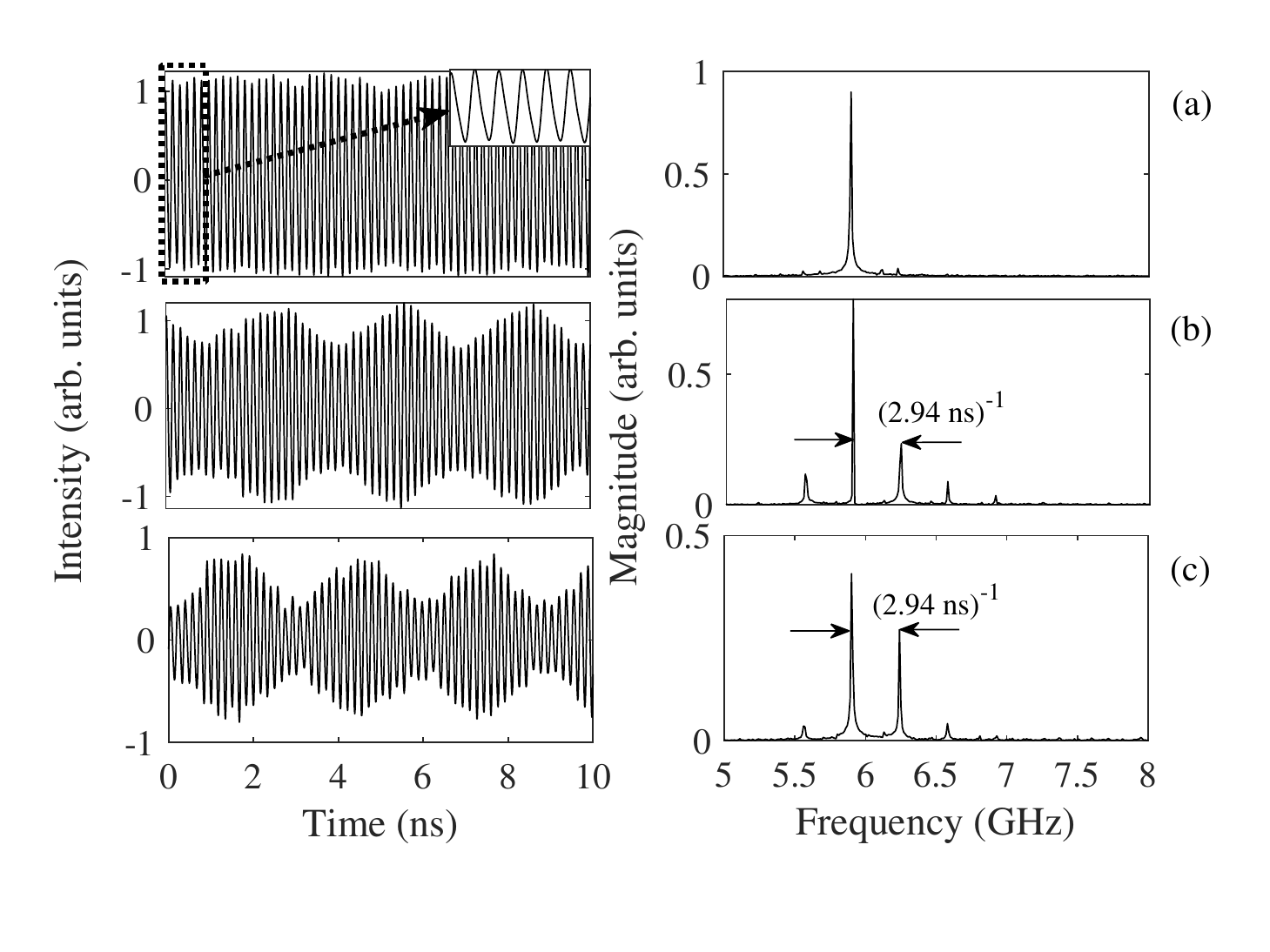}
\caption{Optical intensity (left column) and RF spectra (right column) at 68
mA of pump current for various feedback levels: (a) $\protect\eta =0.56$,
(b) $\protect\eta =0.58$, (c) $\protect\eta =0.59$.}
\label{expfeed}
\end{figure}

The free running laser operates at 68 mA pump current, providing 4 mW output
power. For low feedback strength $\eta $, the laser output remains stable
(apart from noise). Using LP as a variable optical attenuator in the optical
path, the feedback level $\eta $ was increased until cw operation was lost
, and a Hopf bifurcation appears. The latter leads to sustained
relaxation oscillations.\ A regular, nearly-sinusoidal 5.9 GHz\textsl{\ }%
oscillation is obtained at $\eta =0.56$ (see Fig.~\ref{expfeed}(a)). The experimental feedback strength $\eta$ is in arbitrary units, but is proportional to the current fed back into the injection terminals. When $\eta = 0.56$, the fed back current is approximately 15\% the injection current.

As the level of feedback $\eta $ is further increased, the generation
of sidebands in the RF spectra, spaced at approximately $\sim$0.34 GHz (slow period $\sim$2.94 ns), was observed indicating a new bifurcation
transition. This bifurcation arises at $\eta =0.58$ as the output displays
quasiperiodic intensity traces slightly affected by the noise in the system
(Fig.~\ref{expfeed}(b)). Further increase of the feedback sufficiently
affects the amplitude of the slow envelope and its shape.\ The slow period,
however, remains constant for the whole feedback range. It is worth noting
here that the slow period of the quasiperiodic oscillations $\sim$2.94 ns
is nearly three times smaller than that of the $\sim$8.57 ns as
estimated delay time.

\begin{figure}[th]
\includegraphics[width=\linewidth]{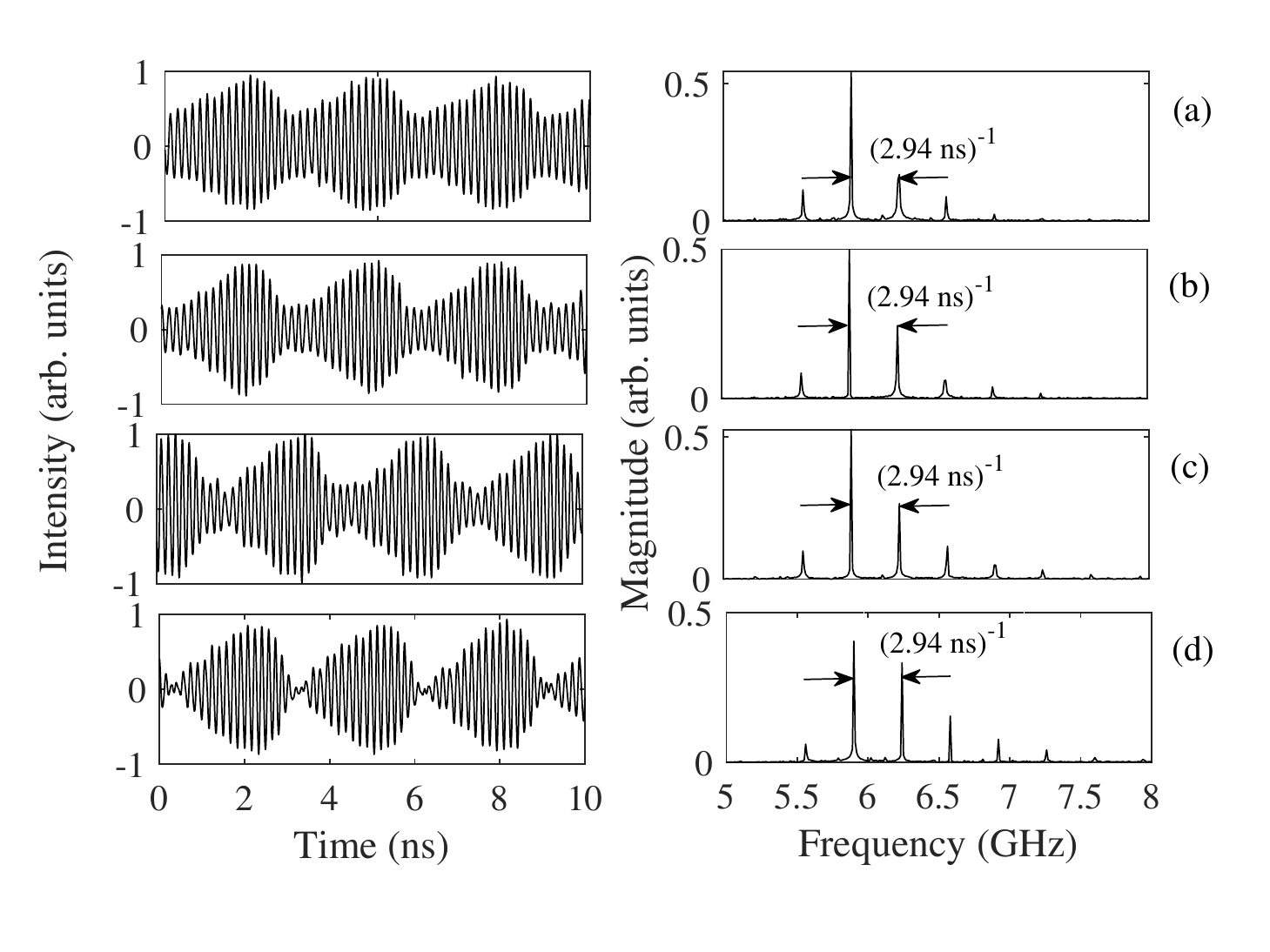}
\caption{Optical intensities (left) and RF spectra (right) for a range of
pump currents $J$ and feedback levels: (a) $J$ = 64~mA, $\protect\eta  =
0.56$, (b) $J$ = 65~mA, $\protect\eta  = 0.57$, (c) $J$ = 66~mA, $\protect%
\eta = 0.58$, (d) $J$ = 67~mA, $\protect\eta = 0.59$.}
\label{expcurr}
\end{figure}

The effective time scales measured in the experiment result from $1:17$
resonance between the low $\sim$0.34 GHz and $large$ $\sim$5.9 GHz frequencies of the quasiperiodic
oscillations. In order to verify the model assumptions about the high-order resonant effect, we have also varied the pump current which led to noticeable changes in the shape and amplitude of the quasiperiodic oscillations. However, as shown in Fig.~\ref{expcurr}, the
fast and slow fundamental frequencies were not affected as they remained
fixed (to $\sim$5.9 GHz and $\sim$0.34 GHz, respectively) for the whole
range of the control parameters.

\section{Conclusion}

In this paper, we considered a semiconductor laser with delayed
optoelectronic feedback. We prove, analytically, and demonstrate, numerically %
and experimentally, that resonant locking between two fundamental laser
frequencies, namely, $f_{RO}$ and $f_{delay}$, allows their
ratio to remain constant despite growing amplitude oscillations. Understanding of SLs with optoelectronic feedback is a relatively undeveloped field compared with purely optical feedback; however,
understanding these effects is important from the viewpoints of stabilizing desired dynamics, gaining insight into undesirable effects of 
feedback, as well as to explore novel nonlinear dynamical effects.

\begin{acknowledgments}
A.V.K, E.A.V. and T.E. acknowledge the Government of Russian Federation (Grant 08-08).
A.L, D.C, and S.I. acknowledge the financial support
of the Conseil R\'{e}gional Grand Est and of the Fond Europ\'{e}en de D\'{e}veloppement R\'{e}gional (FEDER).
\end{acknowledgments}

\bibliography{fundamental-resonance39}

\appendix
\section{Asymptotic methods based on the large delay limit \label{MS}%
}

Asymptotic methods for delay differential equations exhibiting a large
delay take advantage of the distinct time scales in the physical problem.\
In particular, Hopf bifurcation instabilities have been studied in detail,
and the predictions of their amplitude equations have been successfully
tested on several case studies. It is worth emphasizing that there exist two
distinct limits that provide valuable information in their domains of
validity.\ We illustrate these different approaches by considering Minorsky
equation for a weakly damped and weakly nonlinear oscillator \cite{pinney}:%
\begin{equation}
y^{\prime \prime }+\varepsilon y^{\prime }+y=-\varepsilon dy^{\prime }(t-\tau
)+\varepsilon cy^{\prime^3 }(t-\tau ), \label{MS1}
\end{equation}%
where $\varepsilon $  is small and $\tau $ is large.\
Assuming discrete values $\tau =(1+2n)\pi $, where $n$ is a
large integer, allows the derivation of an amplitude equation in \ its
simplest mathematical form. Specifically, we scale the delay $\tau $ 
with respect to $\varepsilon $ as $\tau =\varepsilon ^{-1}\tau _{1}$, and find that the first Hopf bifurcation of Eq.~(\ref{MS1}) leads to
the solution \cite{erneux}:
\begin{equation}
y=A(s)\exp (it)+c.c.+\mathcal{O}(\varepsilon ),  \label{MS2}
\end{equation}%
where the complex amplitude $A$ depends on the slow time
variable $s=\varepsilon t$. It satisfies the slow time equation%
\begin{equation}
A^{\prime }=\frac{1}{2}\left[ -A+dA(s-s_{1})-3cA^{3}(s-s_{1})\right],
\label{MS3}
\end{equation}%
where prime now means differentiation with respect to time $s$, and $s_{1}\equiv \varepsilon \tau _{1}=\mathcal{O}(1)$ or larger. This
equation is analyzed in \cite{erneux} and reveals a cascade of primary and
secondary bifurcations.

If we now analyze the stability of the zero solution of Eq.~(\ref{MS3}%
) using $d$  as the bifurcation parameter, we observe that the first
Hopf bifurcation point $d_{H}\rightarrow 1$ as $s_{1}\rightarrow
\infty .\ $ The nature of the bifurcation corresponds to an uniform
instability according to Ref.\ \cite{giacomelli}.\ Introducing the small
parameter $\delta \equiv s_{1}^{-1}\ll1,$ and expanding $d$ 
as $d=1+\delta ^{2}d_{2}+...$, we may construct a small amplitude
solution of the form \cite{mathias,giacomelli} %
\begin{equation}
A=\delta u(x,\nu )+\mathcal{O}(\delta ^{2})  \label{MS4}
\end{equation}%
where $x\equiv \varepsilon (1+\varepsilon /a+\varepsilon ^{2}/a^{2})t$%
and $\nu \equiv \delta ^{3}t$ are called pseudo-space and
pseudo-time, respectively \cite{giacomelli}. The function $u$%
satisfies the Ginzburg-Landau (GL) equation %
\begin{equation}
u_{\nu }=2u_{xx}+b_{2}u-3cu^{3},  \label{MS5}
\end{equation}%
\begin{equation}
u(x-1)=u(x).  \label{MS6}
\end{equation}%
An obvious question is how to relate the small parameters $\delta \ $%
and $\varepsilon .$ By considering $\delta =\varepsilon ,$%
or equivalently, $\tau =\varepsilon ^{-2}\tau _{2},$ and
seeking a solution of Eq.~(\ref{MS1}) in powers of $\varepsilon $
leads to Eqs.~(\ref{MS5}) and (\ref{MS6})\ \cite{erneux}.

In summary, the limit $\tau =\varepsilon ^{-1}\tau _{1}$ is
leading to a slow time DDE for $\mathcal{O}(1)$ amplitude multi-periodic
solutions.\ On the other hand the limit, $\tau =\varepsilon ^{-2}\tau _{2}$ 
is leading to a GL\ equation for small amplitude solutions.\ The
latter allows to relate our DDE problem to spatially extended systems \cite
{tito,grigorieva}. Here, we consider the first limit because it
quantitatively describes the instabilities observed numerically, and allows us
to analyze high-order locking phenomena, manifested by the resonance between the
multiple timescales in the laser subject to the optoelectronic feedback.

\section{The laser amplitude equation and its solutions}
\label{supp}
\subsection{Hopf bifurcations}

We consider the rate equations for a semiconductor laser subject to a
delayed optoelectronic feedback \cite{TE}. In dimensionless form, they are
given by 
\begin{eqnarray}
I^{\prime } &=&2NI,  \label{B1} \\
TN^{\prime } &=&P+\eta I(t-\tau )-N-(1+2N)I,  \label{B2}
\end{eqnarray}%
where $I$ is the intensity of the laser field and $N$ is the carrier
density.\ $P\sim 1$ is the value of the pump parameter above threshold in
the absence of feedback $(\eta =0).$ $T\sim 10^{3}$ is the ratio of the
carrier and photon lifetimes. $\eta <1$ and $\tau \sim 10^{3}$ are the gain
and the delay of the optoelectronic feedback, respectively.\ By introducing
the new variables $x$, $y$, and $s$ defined by 
\begin{equation}
N\equiv \frac{\omega }{2}x,\text{ }I\equiv P(1+y)\text{, and }s\equiv \omega t
\label{R3}
\end{equation}%
where 
\begin{equation}
\omega \equiv \sqrt{\frac{2P}{T}}  \label{R4}
\end{equation}%
is the (angular) RO frequency, we may eliminate the large $T$ parameter
multiplying the left hand side of Eq.~(\ref{B2}). Specifically, we obtain
the following equations for $y$ and $x$%
\begin{eqnarray}
y^{\prime } &=&x(1+y),  \label{R5} \\
x^{\prime } &=&-y+\eta (1+y(s-\theta ))-\varepsilon x\left[ 1+2P(1+y)\right],
\label{R6}
\end{eqnarray}%
where 
\begin{equation}
\varepsilon \equiv \frac{\omega }{2P}<<1\text{, and }\theta =\omega \tau .
\label{R7}
\end{equation}

\begin{figure}[tb]
\centering
\includegraphics[width=\linewidth]{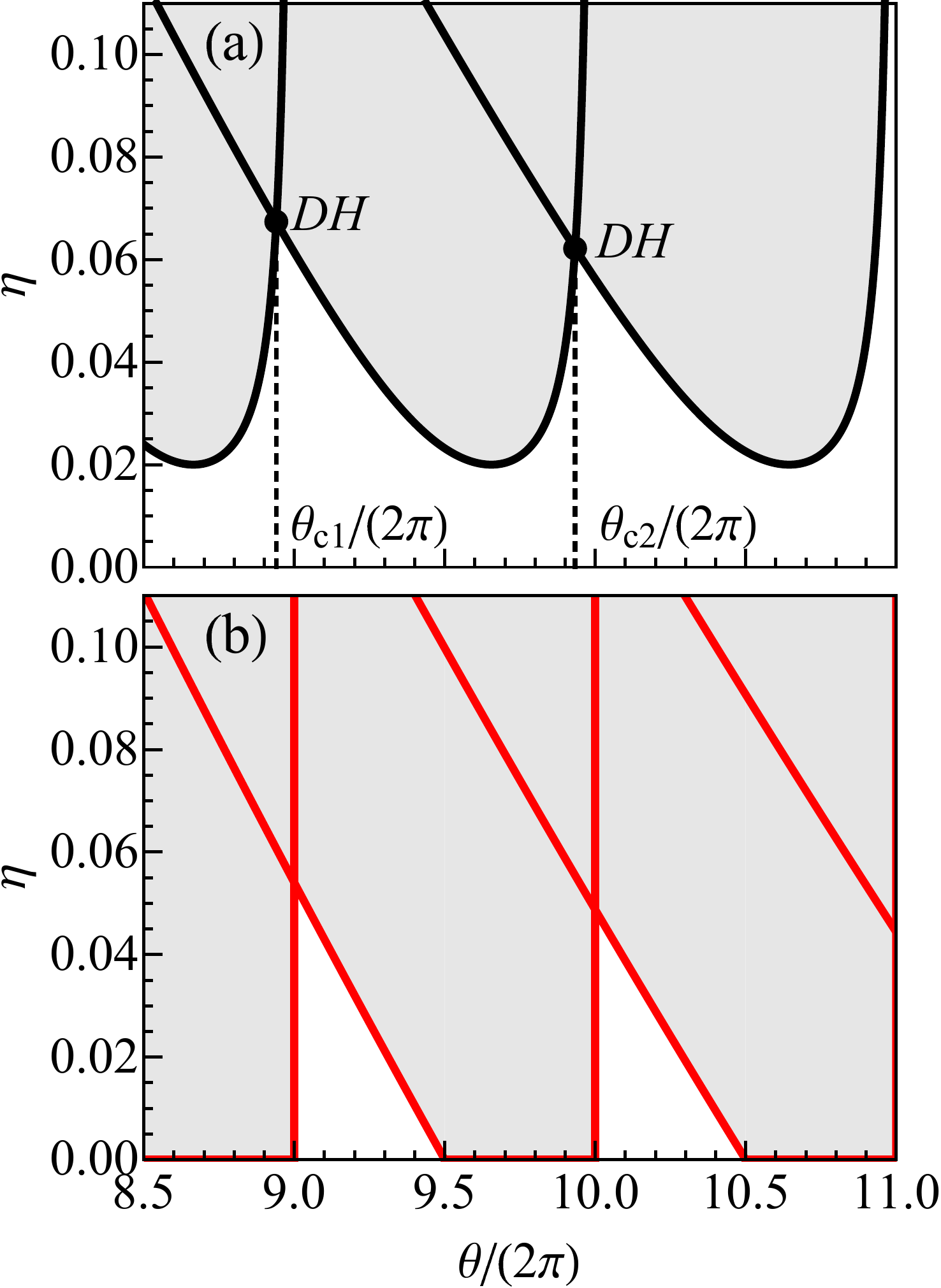}
\caption{(a) Primary Hopf bifurcation lines in the interval $8.5<\protect%
\theta /(2\protect\pi )<11$ as we increase $\protect\eta $ from zero $(%
\protect\varepsilon =0.01$ and $P=0.5).$ At double Hopf points $DH$ (black dots), $\protect\theta 
_{c1}=8.94\times 2\protect\pi$ and $\protect\theta 
_{c2}=9.93\times 2\protect\pi$, two distinct bifurcation lines are
crossing. (b) Primary Hopf bifurcation lines if $\protect\varepsilon =0.$
They are given by (\protect\ref{R12d}) and (\protect
\ref{R12e}) with $n=9,$ $10,$ $11$, and by (\protect\ref{R12c}). The shaded (white) areas in (a) and (b) denote the zones of unstable (stable) steady state solutions.
}
\label{hopfrostab1}
\end{figure}

The non-zero intensity steady state is 
\begin{equation}
(x,y)=(0,\frac{\eta }{1-\eta }).  \label{R50}
\end{equation}%
From the linearized equations, we determine the characteristic equation for
the growth rate $\lambda $.\ We find 
\begin{equation}
\lambda ^{2}+\varepsilon \lambda (1+\frac{2P}{1-\eta })+1-\frac{\eta }{%
1-\eta }(\exp (-\lambda \theta )-1)=0.  \label{R50b}
\end{equation}%
The stability domains in the $(\eta ,\theta )$ parameter space are bounded
by Hopf bifurcation lines. Introducing $\lambda =i\sigma $ into Eq.~(\ref%
{R50b}), we obtain the Hopf conditions relating $\eta $ \ and $\sigma .$
They are given by 
\begin{eqnarray}
-\sigma ^{2}+1-\frac{\eta }{1-\eta }(\cos (\sigma \theta )-1) &=&0,
\label{R9} \\
\sigma \varepsilon (1+\frac{2P}{1-\eta })+\frac{\eta }{1-\eta }\sin (\sigma
\theta ) &=&0.  \label{R10}
\end{eqnarray}%
Figure \ref{hopfrostab1} shows the stability domains for $\varepsilon =0.01$
and for $\varepsilon =0$ $(8.5<\theta /(2\pi )<11)$. As $\varepsilon
\rightarrow 0,$ the Hopf stability boundaries are shrinking to straight
lines. An analysis of Eq.~(\ref{R50b}) with $\varepsilon =0$ and $\eta
\rightarrow 0$ leads to the stability condition 
\begin{equation}
\sin (\theta )>0\text{ }(\varepsilon =0,\eta \rightarrow 0).  \label{R10a}
\end{equation}%
This explains the sequential change of stability along the $\eta =0$ axis in
Fig.~\ref{hopfrostab1}(b).

If $\varepsilon =0,$ Eqs.~(\ref{R9}) and (\ref{R10}) are easily solved.\ We
find three families of solutions given by

\begin{align}
\text{(1): }&\sigma \theta =2n\pi \text{ }(n=1,...)\text{, and }\sigma =1,
\label{R11} \\
\text{(2): }&\sigma \theta =(2n+1)\pi \text{ }(n=0,\text{ }1,...)\nonumber\\
&\text{ and }\sigma =\sqrt{1+\frac{2\eta }{1-\eta }},  \label{R12a} \\
\text{(3): }&\eta =0 \text{, and }\sigma =1.  \label{R12c}
\end{align}%
The three cases provide the vertical lines in Fig.~\ref{hopfrostab1}(b):
\begin{equation}
\text{(1): }\theta =2n\pi \text{ }(n=1,...),  \label{R12d}
\end{equation}%
the lines 
\begin{equation}
\text{(2): }\theta =\frac{(2n+1)\pi }{\sqrt{1+\frac{2\eta }{1-\eta }}}\text{ }%
(n=0,\text{ }1,...),  \label{R12e}
\end{equation}%
and the horizontal line (\ref{R12c}).

\subsection{Perturbation analysis}

We next consider values of $\theta $ close to%
\begin{equation}
\theta _{n}\equiv 2n\pi  \label{R51a}
\end{equation}%
and wonder if a secondary bifurcation is possible$.$ Numerical simulations
of Eqs.~(\ref{R5}) and (\ref{R6}) suggest that such bifurcation appears at a
value of $\eta $ satisfying the scaling law $\eta \sim n^{-1}$ for large $n$%
.\ It motivates a weakly nonlinear analysis where \ 
\begin{equation}
\delta \equiv 1/(2n)  \label{R52}
\end{equation}%
will be considered as a small parameter. To facilitate the algebra, it will
be convenient to eliminate $x$ and formulate a second order delay
differential equation for $u\equiv \ln (1+y)$\ only. From Eqs.~(\ref{R5})
and (\ref{R6}), we find that $u$ satisfies 
\begin{align}
u^{\prime \prime }=&1-\exp (u)+\eta \exp (u(s-\theta ))\nonumber\\
&-\varepsilon u^{\prime }(1+2P\exp (u)).  \label{R13}
\end{align}%
We are now ready to start our analysis.\ We introduce the new control
parameters $b=\mathcal{O}(1)$ and $c=\mathcal{O}(1)$ defined as 
\begin{equation}
b\equiv \eta \delta ^{-1},\text{ }c\equiv \varepsilon \delta ^{-1},
\label{R14}
\end{equation}%
and seek a solution depending on two distinct time variables of the form 
\begin{equation}
u=\delta ^{1/2}u_{1}(s,r)+\delta u_{2}(s,r)+...,  \label{R16}
\end{equation}%
where $r\equiv \delta s$ is defined as a slow time variable. The $\delta
^{1/2}$ power series in (\ref{R16}) and the scaling of $\eta $, and $r$
result from the fact that the desired amplitude equation only appears at the
third order of the perturbation analysis. The assumption of two independent
time scales implies the chain rule 
\begin{equation}
u^{\prime \prime }=u_{ss}+2\delta u_{sr}+\delta ^{2}u_{rr},  \label{R17}
\end{equation}%
where the subscripts $s$ and $r$ mean partial derivatives with respect to $s$
and $r$. We also note that 
\begin{equation}
u(s-\theta )=u(s-\theta ,r-\delta \theta ).  \label{R18}
\end{equation}%
Introducing (\ref{R14})--(\ref{R18}) into Eq.~(\ref{R13}) and equating to
zero the coefficients of each power $\delta ^{1/2}$ lead to a sequence of
linear problems for the unknown functions $u_{1}$, $u_{2}$, and $u_{3}$. They
are given by
\begin{widetext}
\begin{align}
\mathcal{O}(\delta ^{1/2}): u_{1ss}+u_{1}&=0,  \label{R19} \\
\mathcal{O}(\delta ): u_{2ss}+u_{2}&=b-\frac{u_{1}^{2}}{2},  \label{R20} \\
\mathcal{O}(\delta ^{3/2}): u_{3ss}+u_{3}&=\left[ 
\begin{array}{c}
bu_{1}(s-\theta ,r-\delta \theta )-u_{1}u_{2}-\frac{u_{1}^{3}}{6}-2u_{1sr}
\\ 
-cu_{1s}(1+2P\exp (u))%
\end{array}%
\right].  \label{R21}
\end{align}
\end{widetext}
The solution of Eqs.~(\ref{R19}) and (\ref{R20}) are 
\begin{align}
u_{1} =&A(r)\exp (is)+c.c.,  \label{R22} \\
u_{2} =&B(r)\exp (is)+c.c.\nonumber\\
&+b-AA^{\ast }+\frac{1}{6}A^{2}\exp (2is)+c.c.
\label{R23}
\end{align}%
where $A(r)$ and $B(r)$ are two unknown amplitudes. In order to determine an
equation for $A(r)$, we consider Eq.~(\ref{R21}) and apply a solvability
condition. We cannot neglect $\delta \theta $ in $u_{1}(s-\theta ,r-\delta
\theta )$ because we assume $\theta $ close to $\theta _{n}=2n\pi $ and
therefore $\delta \theta \sim \pi $ is an $\mathcal{O}(1)$ quantity.\ The solvability
condition requires that there are no terms of the form $\exp (\pm is)$ in
the right hand side of Eq.~(\ref{R21}).\ This condition leads to a delay
differential equation for $A$ given by 
\begin{align}
2i\frac{dA}{dr}=&\frac{1}{3}A^{2}A^{\ast }-Ab+bA(r-\delta \theta )\exp
(-i\theta )\nonumber\\
&-icA(1+2P).  \label{R24}
\end{align}%
Introducing $A=R\exp (i\phi )$ into Eq.~(\ref{R24}), we obtain from the real
and imaginary parts, two coupled equations for $R$ and $\phi $ 
\begin{align}
2R^{\prime } =&bR(r-\delta \theta )\sin (-\theta +\phi (r-\delta \theta
)-\phi )\nonumber\\
&-icR(1+2P),  \label{R25} \\
2\phi ^{\prime } =&-b\frac{R(r-\delta \theta )}{R}\cos
(-\theta +\phi (r-\delta \theta )-\phi )\nonumber\\
&-\frac{1}{3}R^{2}+b.  \label{R26}
\end{align}

\subsection{Primary and secondary bifurcation ($\protect\varepsilon =0)$}

For mathematically clarity, we now propose an analysis of Eq.~(\ref{R25})
and (\ref{R26}) with $c=0$. Time periodic solutions of the original laser
equations (\ref{R5}) and (\ref{R6}) correspond to solutions of Eqs.~(\ref%
{R25}) and (\ref{R26}) of the form 
\begin{equation}
R=\text{const, and }\phi =\nu r,  \label{R27}
\end{equation}%
where $\nu $ is the frequency correction. Inserting (\ref{R27}) into Eqs.~(%
\ref{R25}) and (\ref{R26}), we obtain the conditions 
\begin{align}
\sin (\theta +\nu \delta \theta )=0,\nonumber\\
\text{ and }\nonumber\\
2\nu =-\frac{1}{3}R^{2}+b-b\cos (\theta +\nu \delta \theta ).  \label{R28}
\end{align}%
We analyze these equations for $\theta $ close to $\theta _{n}$ by
introducing 
\begin{equation}
\theta =\theta _{n}+\Theta,  \label{R29}
\end{equation}%
where $0\leq |\Theta |<2\pi $. The possible solutions of Eq.~(\ref{R28})
then are 
\begin{widetext}
\begin{align}
\text{(1): }&\theta _{n}+\Theta +\nu (\delta \theta _{n}+\delta \Theta )=2n\pi
+\Theta +\nu \pi +\mathcal{O}(\delta \Theta )=m2\pi ,\text{ }  \label{R30} \\
&R^{2} =-6\nu \geq 0,  \label{R31} \\
\text{(2): }&\theta _{n}+\Theta +\nu (\delta \theta _{n}+\delta \Theta )=2n\pi
+\Theta +\nu \pi +\mathcal{O}(\delta \Theta )=(2m+1)\pi ,  \label{R32} \\
&R^{2} =-6\nu +6b\geq 0.  \label{R33}
\end{align}%
\end{widetext}
The first case matches the stable Hopf bifurcation points at $\theta =\theta
_{n}$ if $\nu =0$ and $\Theta =0\ (m=n).\ $If $\Theta >0,$ the first
solution is for $m=n$ and $\nu =-\Theta /\pi $ which provides 
\begin{equation}
R^{2}=6\Theta /\pi >0.  \label{R33b}
\end{equation}%
If $\Theta <0,$ the first solution is for $m=n-1$ and 
\begin{equation}
\nu =-(\pi +\Theta )/\pi,  \label{R33c}
\end{equation}
which then leads, using (\ref{R33}), to 
\begin{equation}
R^{2}=6\left[ (\pi +\Theta )/\pi +b\right] >0.  \label{R34}
\end{equation}%
In order to explore the onset of a bifurcation point from the periodic
solution, we consider the linearized equations from Eqs.\ (\ref{R25}) and (%
\ref{R26}). The characteristic equation for the growth rate $\lambda $ is
obtained from the condition 
\begin{widetext}
\begin{equation}
\left\vert 
\begin{tabular}{ll}
$\left[ 
\begin{array}{c}
-b\sin (\theta +\nu \delta \theta ) \\ 
\times \exp (-\lambda \delta \theta ) \\ 
-2\lambda%
\end{array}%
\right] $ & $\left[ 
\begin{array}{c}
bR\cos (\theta +\nu \delta \theta ) \\ 
\times (\exp (-\lambda \delta \theta )-1)%
\end{array}%
\right] $ \\ 
$\left[ 
\begin{array}{c}
-\frac{2}{3}R \\ 
-\frac{b}{R}\cos (\theta +\nu \delta \theta ) \\ 
\times (\exp (-\lambda \delta \theta )-1)%
\end{array}%
\right] $ & $\left[ 
\begin{array}{c}
-b\frac{R(r-\delta \theta )}{R}\sin (\theta +\gamma \delta \theta ) \\ 
\times (\exp (-\lambda \delta \theta )-1) \\ 
-2\lambda%
\end{array}%
\right] $%
\end{tabular}%
\right\vert =0.  \label{R54}
\end{equation}%
\end{widetext}
The coefficients in (\ref{R54}) simplify, if we take into account the fact
that $\sin (\theta +\nu \delta \theta )=0$ and $\cos (\theta +\nu \delta
\theta )=-1$ for the Hopf bifurcation appearing if $\Theta <0.$ It leads to
the following equation for $\lambda $ 
\begin{align}
0=&4\lambda ^{2}-\frac{2}{3}bR^{2}(\exp (-\lambda \delta \theta )-1)\nonumber\\
&+b^{2}(\exp
(-\lambda \delta \theta )-1)^{2}.  \label{R37}
\end{align}%
We are interested to find if a Hopf bifurcation for the slow time equations (%
\ref{R25}) and (\ref{R26}) is possible. Recall that it will correspond to a
secondary bifurcation of the original laser equations (\ref{R5}) and (\ref%
{R6}).\ To this end we introduce $\lambda =i\mu $ into (\ref{R37}) and
determine from the real and imaginary parts two conditions 
\begin{align}
0=&-4\mu ^{2}-\frac{2}{3}bR^{2}(\cos (\mu \delta \theta )-1)\nonumber\\
&+b^{2}((\cos (\mu\delta \theta )-1)^{2}-\sin ^{2}(\mu \delta \theta )),  \label{R55} \\
0=&2b\sin (\mu \delta \theta )\left( \frac{1}{3}R^{2}-b(\cos (\mu \lambda
\theta )-1)\right).  \label{R56}
\end{align}%
From Eq.~(\ref{R56}), a first possibility is given by the condition $\sin
(\mu \delta \theta )=0.\ $It implies 
\begin{align}
\text{(1): }&\mu \delta \theta =2k\pi \text{ }(k=1,2,...),  \label{R57} \\
\text{(2): }&\mu \delta \theta =(2k+1)\pi \text{ }(k=0,1,2...).  \label{R57b}
\end{align}%
(\ref{R57}) is not possible because $\cos (\mu \delta \theta )=1$, and Eq.~(%
\ref{R57}) can be satisfied only if $\mu =0.$ We next consider (\ref{R57b})
and take the lowest value of $\mu $ $(k=0)$ given by%
\begin{equation}
\mu =\pi /(\delta \theta ).  \label{R57a}
\end{equation}%
From (\ref{R55}), we then obtain 
\begin{equation}
4\left( -\pi ^{2}/(\delta \theta )^{2}+\frac{1}{3}bR^{2}+b^{2}\right) =0.
\label{R58}
\end{equation}%
Using (\ref{R29}), we have 
\begin{equation}
\delta \theta =\pi +\Theta /(2n).  \label{R60}
\end{equation}%
Substituting (\ref{R34}) into (\ref{R58}), we solve for $b=b_{SB}$ and find 
\begin{equation}
b_{SB}=\frac{1}{3}\left[ -\frac{\pi +\Theta }{\pi }+\sqrt{(\frac{\pi +\Theta 
}{\pi })^{2}+3}\right] .  \label{R59}
\end{equation}%
This secondary bifurcation is characterized by two frequencies namely, the
frequency of the basic periodic solution 
\begin{equation}
\omega _{1}=1+\delta \nu,  \label{R59a}
\end{equation}
and the slow time frequency 
\begin{equation}
\omega _{2}=\delta \mu .  \label{R59b}
\end{equation}%
Using (\ref{R33c}) for $\nu $, (\ref{R52}) for $\delta ,$ (\ref{R57a}) for $%
\mu $, and (\ref{R60}) for $\delta \theta ,$ we obtain from (\ref{R59a}) and
(\ref{R59b}) 
\begin{align}
\omega _{1} &=1-\frac{\pi +\Theta }{2n\pi },  \label{R61} \\
\omega _{2} &=\frac{\delta \pi }{\delta \theta }=\frac{\pi }{2n\delta
\theta }=\frac{1}{2n}(1-\frac{\Theta }{2n\pi }+\mathcal{O}((2n)^{-2}).  \label{R62}
\end{align}%
The ratio of the frequencies clearly verifies the ratio%
\begin{equation}
\frac{\omega _{2}}{\omega _{1}}=\frac{1}{2n}+\mathcal{O}((2n)^{-2}),  \label{R63}
\end{equation}%
and parameter $\Theta $ does not appear in the leading
approximation.\ 

\end{document}